\documentclass[12pt,twoside,fleqn]{article}

\usepackage{a4wide,cite}
\usepackage{amsmath,amssymb} 
\usepackage{epsfig,rotating}

\usepackage[dvips]{pstcol}

\usepackage{axodraw}

\numberwithin{equation}{section}
\allowdisplaybreaks[4]

\def \be{\begin{equation}}
\def \ee{\end{equation}}

\def \nn{\nonumber}
\newcommand{\MS}{\ensuremath{\overline{\text{MS}}}}

\begin{document}

\begin{titlepage}

\flushright{CERN-PH-TH/2006-074\\ SLAC-PUB-11841}
\vspace{2cm}

\centerline{\Large\bf NNLL QCD  Contribution of the Electromagnetic}
\vspace{2mm}
\centerline{\boldmath \Large\bf Dipole Operator to 
$\Gamma(\bar{B}\to X_s\gamma$)}
\vspace{2.5cm}

\begin{center}
{\bf H.M.~Asatrian, A.~Hovhannisyan, and V.~Poghosyan}\\
{\it Yerevan Physics Institute, 375036 Yerevan, Armenia}

{\bf T.~Ewerth and C.~Greub}\\
{\it Inst. for Theoretical Physics, Univ. Berne, CH-3012 Berne, Switzerland}

{\bf T.~Hurth}\footnote{Heisenberg Fellow}\\
{\it CERN, Dept.\ of Physics, Theory Unit, CH-1211 Geneva 23, Switzerland}\\ 
{\it SLAC, Stanford University, Stanford, CA 94309, USA}
\end{center}
\medskip

\vspace{2cm}

\begin{abstract}
We present an independent calculation of that part of the $O(\alpha_s^2)$ 
contribution to the decay width $\Gamma(\bar{B} \to X_s \gamma)$
which arises from the self-interference term
of the electromagnetic dipole operator ${\cal O}_7$.  
Using a different method, we find complete agreement with a previous
calculation.
This NNLL contribution  is an important ingredient for the complete
NNLL prediction of $\Gamma(\bar{B} \to X_s\gamma)$
which will resolve the charm quark mass ambiguity appearing at NLL accuracy.
\end{abstract}

\end{titlepage}

\section{Introduction}

Rare $B$ decays as flavor changing neutral currents (FCNCs)
are  loop-induced processes 
and highly sensitive  probes for new degrees of freedom  beyond the SM.
But this  indirect search for new physics 
signatures  within flavor physics takes place today in complete darkness, 
given that  we have no direct evidence of new particles beyond the SM.  
However, the day the existence of new degrees of freedom  is established  by 
the LHC, specifically when the mass scale of the new physics is fixed, 
the searches for anomalous  phenomena  in the flavor sector will  
become mandatory. In this context, 
the measurement of  theoretically  clean rare  decays, even when found 
to be SM-like, will lead to  important and  valuable  information  on
the structure of the new-physics models and to complementary
information to the LHC collider data. 

Because physics effects beyond the SM in flavor observables 
seem to be rather small,
as recent experimental data indicate, the focus on theoretically 
clean observables within the indirect search for new physics is crucial
in order to detect specific patterns and to distinguish between 
various new physics 
scenarios.  Inclusive rare $B$ decays belong to this class of observables
 because they  are  dominated by perturbative contributions and  
non-perturbative effects are generally suppressed by $\Lambda_{\rm QCD}^2/m_b^2$
within the heavy mass expansion (for a review see \cite{Hurth:2003vb}).

The inclusive $b \rightarrow s \gamma$ mode is still the most prominent
rare decay. 
The stringent bounds obtained from this mode
on various non-standard scenarios (see
e.g.~\cite{Bertolini:1986tg,Degrassi:2000qf,Carena:2000uj,D'Ambrosio:2002ex,MFVnew,
Borzumati:1999qt,Besmer:2001cj,Ciuchini:2002uv})
indicate the importance of this specific FCNC observable in discriminating
new-physics models.  
It has been measured by several independent experiments
\cite{belle1,cleobsg,babar1,babar2,belle2},  
and the present experimental accuracy is already below  the $10 \%$ level.
In the HFAG world  average of these measurements,  a common shape function for
the extrapolation to low energies was used;
for a  photon energy cut $E_\gamma > 1.6$ GeV the average is given by \cite{HFAG}:
\begin{eqnarray}
\mbox{BR}[\bar B \to X_s \gamma]  = \left(3.55\pm0.24^{~+~0.09}_{~-~0.10}\pm0.03\right) \times  10^{-4} \,,
\label{world} 
\end{eqnarray}
where the errors are combined statistical and systematic, 
the systematic ones due to shape function, and due to the $d\gamma$ 
contamination. 
In the near future, more precise data on this mode are expected from the 
(upgraded) $B$ factories and from the planned Super-$B$ factories
(see \cite{Superbelle,Superbabar,Linearsuperb}). 

Clearly, the theoretical uncertainty should be reduced
accordingly. After completion of the QCD NLL
\cite{Adel,GHW,Mikolaj,GH,Burasnew,Paolonew,Buras:2002tp,Asatrian:2004et,Ali:1990tj,Pott:1995if}. 
and the electroweak two-loop \cite{Czarnecki:1998tn,Kagan:1998ym,Baranowski:1999tq,Gambino:2000fz}
computations, it was generally believed that the theoretical uncertainty of the branching ratio is below $10\%$. 
However, as first pointed out in \cite{GambinoMisiak},
there is an additional uncertainty in the NLL prediction,  
which is related to the renormalization scheme of the charm-quark mass. 
In a recent theoretical update of the NLL prediction of the branching ratio of 
$\bar B \rightarrow X_s \gamma$,
the uncertainty related to the definition of $m_c$ was taken into account
by varying $m_c/m_b$ in the conservative range $0.18 \le m_c/m_b \le 0.31$, 
which covers both  the pole mass (with its numerical error) value and 
the running mass 
$\bar{m}_c(\mu_c)$ value with $\mu_c \in [m_c,m_b]$;
for $E_\gamma > 1.6$ GeV one then finds \cite{Hurth:2003dk}
\be
\label{hurth_lunghi}
\mbox{BR}[\bar{B} \to X_s \gamma] = (3.52 \pm 0.32|_{m_c/m_b} \pm
0.02|_{\rm{CKM}} \pm 0.24|_{\rm{param.}} \pm 0.14|_{\rm{scale}}) \times 10^{-4} \, .
\ee
The only way to resolve this scheme ambiguity in a satisfactory way is 
to perform a systematic NNLL calculation,  which according to a recent
investigation \cite{New} is expected to reduce this
uncertainty by a factor 2.


Parts of the  three principal calculational 
steps leading to the NNLL result within the
effective field theory approach  are already done:
{\bf (a)} The full SM theory has to be matched 
with the effective theory at the scale $\mu=\mu_W$, where
$\mu_W$ denotes a scale of order $m_W$ or $m_t$. 
The Wilson coefficients 
$C_i(\mu_W)$ only pick up small QCD corrections,
which can be calculated in fixed-order perturbation theory. 
In the NNLL  program, the matching has to be worked out at  
order $\alpha_s^2$. The matching calculation to this precision is already
finished, including the most difficult piece, the three-loop matching 
of the operators ${\cal O}_{7,8}$ \cite{Misiak}.
{\bf (b)} The  evolution of these Wilson 
coefficients from  $\mu=\mu_W$ down to $\mu = \mu_b$ then has to be performed 
with the help of the renormalization group, where $\mu_b$ is of the 
order of $m_b$.
As the matrix elements of the operators evaluated at the low scale
$\mu_b$ are free of large logarithms, the latter are contained in resummed
form in the Wilson coefficients. For the NNLL  calculation, this RGE step
has to be done  using the anomalous-dimension matrix up  to order $\alpha_s^3$. 
While the three-loop mixing among the four-quark operators 
${\cal O}_i$ $(i=1,\ldots, 6)$ \cite{GorbahnHaisch}
and among the dipole operators ${\cal O}_{7,8}$ \cite{GorbahnHaischMisiak}
are already available, the four-loop mixing of the four-quark into the dipole
operators is still an open issue. {\bf (c)} 
To achieve NNLL  precision, the matrix elements 
$\langle X_s \gamma  |{\cal O}_i (\mu_b)|b \rangle$ have to be calculated 
to order $\alpha_s^2$ precision. This includes also bremsstrahlung corrections. 
In 2003, the ($\alpha_s^2\,N_F$) corrections to the matrix elements of the
operators ${\cal O}_1$,${\cal O}_2$,${\cal O}_7$,${\cal O}_8$ were calculated 
\cite{Bieri}. 
Complete order-$\alpha_s^2$ results are available to the 
$({\cal O}_7,{\cal O}_7)$
contribution to the decay width \cite{Blokland:2005uk}.
Recently, also order $\alpha_s^2$ terms
to the photon energy spectrum (away from the endpoint $E_\gamma^{\rm{max}}$)
were worked out for the operator ${\cal O}_7$ \cite{Melnikov:2005bx}.

In this paper we present our calculation of the $O(\alpha_s^2)$ contribution 
to the partonic $b \rightarrow X_s^{\rm parton} \gamma$ decay rate due to  
the electromagnetic dipole operator ${\cal O}_7$.  
Contrary to \cite{Blokland:2005uk}, where this self-interference of the 
${\cal O}_7$ operator  was extracted from the imaginary
parts of the corresponding 3-loop diagrams via the optical theorem, we 
calculate the individual cut contributions directly.
The advantage of our approach is that it also allows for the 
calculation of the NNLL interference term of the electromagnetic
operator ${\cal O}_7$ and the chromomagnetic dipole operator ${\cal O}_8$ and also 
of the NNLL self-interference term of the operator ${\cal O}_8$ in a
straightforward manner.
Applying the optical theorem in a manner analogous to the 
$({\cal O}_7,{\cal O}_7)$-interference is not possible in the latter cases,
because some cuts contribute to processes other than the
$b\to X_s^{\rm partonic} \gamma$ decay,  and they cannot be separated. 
We already anticipate here that our results are in complete agreement with the 
ones presented in \cite{Blokland:2005uk}.

The remainder of this paper is organized as follows.
In section 2 we first present our final results, while in section 3  
we discuss in some detail the various steps and ingredients of our
calculation. Section 4 is reserved for our conclusions. 

\section{Results}
\label{sec:2}

Within the low-energy effective theory the partonic $b\to X_s\gamma$ decay rate
can be written as
\be
\Gamma(b\to X_s^{\rm parton}\gamma)_{E_\gamma>E_0} = \frac{G_F^2\alpha_{\rm
    em}m_b^2(\mu)m_b^3}{32\pi^4}
\,|V_{tb}^{}V_{ts}^*|^2\,\sum_{i,j}C_i^{\rm eff}(\mu)\,C_j^{\rm
  eff}(\mu)\,G_{ij}(E_0,\mu)\,,
\label{Gamma_decomp}
\ee
where $m_b$ and $m_b(\mu)$ denote the pole and the running $\MS$ mass of the
$b$ quark, respectively, $C_i^{\rm eff}(\mu)$
the effective Wilson coefficients at the low energy scale, and $E_0$ the
energy cut in the photon spectrum. In the present paper,
we calculate the function $G_{77}(E_0,\mu)$ for $E_0=0$, i.e. without 
a lower cut-off for the photon energy.
The function $G_{77}(E_0,\mu)$
corresponds to the
self-interference of the electromagnetic dipole operator
\be
{\cal O}_7 = \frac{e}{16\pi^2}\,m_b(\mu)\left(\bar s\sigma^{\mu\nu}P_Rb\right)F_{\mu\nu}
\ee
including $\alpha_s^2$ terms as required for NNLL accuracy. When working to
this precision this function can be separated into three
different parts,
\be\label{g77}
G_{77}(0,\mu) = \sum_{n=2}^4G_{77}^{1\to n}(0,\mu) \, ,
\ee
corresponding to the $n$ particles in the final state, namely the $b\to s\gamma$
($n=2$), $b\to s\gamma g$ ($n=3$), $b\to s\gamma gg$ and $b\to s\gamma q\bar
q$ (both $n=4$, $q\in\{u,d,c,s\}$) transitions. The individual contributions
can be written as
\begin{align}
  G_{77}^{1\to 2}(0,\mu) &= 1 + \frac{\alpha_s(\mu)}{4\pi}\,X_{21} +
    \left(\frac{\alpha_s(\mu)}{4\pi}\right)^2X_{22} + O(\alpha_s^3)\nn\\
  G_{77}^{1\to 3}(0,\mu) &= \hspace{.72cm}\frac{\alpha_s(\mu)}{4\pi}\,X_{31} +
    \left(\frac{\alpha_s(\mu)}{4\pi}\right)^2X_{32} + O(\alpha_s^3)\nn\\
  G_{77}^{1\to 4}(0,\mu) &= \hspace{3.05cm}
 \left(\frac{\alpha_s(\mu)}{4\pi}\right)^2X_{42} + O(\alpha_s^3)\,,
\label{twothreefour}
\end{align}
where $\alpha_s(\mu)$ is the running coupling constant in the $\MS$ scheme.
The functions $X_{ij}$ themselves can be naturally divided into two parts,
\be
X_{ij} = X_{ij}^{\rm bare}+X_{ij}^{\rm ct}\,,
\ee
where the superscripts `bare' and `ct' denote the contributions stemming from
the unrenormalized amplitudes and from amplitudes involving at least one
counterterm $Z$ factor, respectively. Working in the Feynman gauge for the
gluon propagator and defining
\be
L_\mu=\ln\left(\frac{\mu}{m_b}\right)
\ee
the contributions of order $\alpha_s$ read (inserting the color factors
in numerical form)
\begin{align}
 X_{21}^{\rm bare} &= -\frac{8}{3\,\epsilon^2}-\frac{1}{\epsilon}\left(8+\frac{32}{3}L_\mu\right) -
   \frac{64}{3}+\frac{4}{9}\pi^2-32 L_\mu-\frac{64}{3}L_\mu^2\nn\\[2mm]
 X_{21}^{\rm ct} &= -\frac{4}{3\,\epsilon}-\frac{20}{3}-\frac{32}{3}L_\mu\nn\\[2mm]
 X_{31}^{\rm bare} &= \frac{8}{3\epsilon^2}+\frac{1}{\epsilon}\left(\frac{28}{3}+\frac{32}{3}L_\mu\right) +
  \frac{316}{9}-\frac{20}{9}\pi^2+\frac{112}{3}L_\mu+\frac{64}{3}L_\mu^2\nn\\[2mm]
 X_{31}^{\rm ct} &= 0\,,
\end{align}
while those of order $\alpha_s^2$ are given by
\begin{align}
 X_{22}^{\rm bare} &= \left(\frac{\mu}{m_b}\right)^{6 \epsilon} \Bigg\{
\frac{ 3.55556}{\epsilon^4}
+\frac{0.444445 \, N_L + 1.77778 \, N_H - 21.5556}{\epsilon^3} \nn\\[2mm]
&\qquad +\frac{2.96296 \, N_L + 4.44444 \, N_H - 58.5668}{\epsilon^2} \nn\\[2mm]
&\qquad +\frac{15.6452 \, N_L + 12.4638 \, N_H - 298.294}{\epsilon} 
  +68.4541 \, N_L + 20.9389 \, N_H - 911.484 \Bigg\}
 \nn\\[4mm]
 X_{22}^{\rm ct} &= \frac{1}{\epsilon^3}\left(\frac{584}{9}-\frac{16}{9}N_F\right) +
   \frac{1}{\epsilon^2}\Bigg\{\frac{1762}{9}+\frac{3104}{9}L_\mu -
   \left(\frac{52}{9}+\frac{64}{9}L_\mu\right) N_F+\frac{4}{3}N_H\Bigg\}\nn\\[2mm]
 &\qquad + \frac{1}{\epsilon}\Bigg\{\frac{5635}{9}-\frac{196}{27}\pi^2+\frac{9172}{9}L_\mu +
   \frac{8512}{9}L_\mu^2\nn\\[2mm]
 &\hspace{2cm} - \left(\frac{382}{27}-\frac{8}{27}\pi^2+\frac{200}{9}L_\mu+\frac{128}{9}L_\mu^2\right) N_F +
   \left(\frac{2}{9}+8L_\mu\right)N_H\Bigg\}\nn\\[2mm]
 &\qquad + \frac{24403}{18}-\frac{1399}{54}\pi^2-\frac{32}{9}\pi^2\ln 2-\frac{4912}{27}\zeta_3 +
   \left(\frac{28162}{9}-\frac{976}{27}\pi^2\right)L_\mu+\frac{24316}{9}L_\mu^2\nn\\[2mm]
 &\qquad + \frac{47872}{27}L_\mu^3 +
   \left(\frac{635}{27}-\frac{41}{9}\pi^2+\frac{4}{3}L_\mu+24L_\mu^2\right)N_H\nn\\[2mm]
 &\qquad - \left(\frac{583}{27}-\frac{17}{9}\pi^2-\frac{128}{27}\zeta_3 +
   \left(\frac{1076}{27}-\frac{32}{27}\pi^2\right)L_\mu+\frac{344}{9}L_\mu^2+\frac{512}{27}L_\mu^3\right)N_F\nn
\end{align}

\begin{align}
 X_{32}^{\rm bare} &= \left(\frac{\mu}{m_b}\right)^{6 \epsilon} \Bigg\{
-\frac{11.1111}{\epsilon^4}
-\frac{31.1111}{\epsilon^3} 
-\frac{89.9399}{\epsilon^2} 
-\frac{262.232}{\epsilon} -1335.13 \Bigg\} \nn\\[4mm]
X_{32}^{\rm ct} &=
-\frac{1}{\epsilon^3}\left(\frac{584}{9}-\frac{16}{9}N_L\right)
   -\frac{1}{\epsilon^2}\Bigg\{236+\frac{3104}{9}L_\mu -
\left(\frac{56}{9}+\frac{64}{9}L_\mu\right)N_L +
   \frac{32}{9}L_\mu N_H\Bigg\}\nn\\[2mm]
 &\qquad
-\frac{1}{\epsilon}\Bigg\{\frac{28348}{27}-\frac{1364}{27}\pi^2+\frac{11344}{9}L_\mu
+  \frac{8512}{9}L_\mu^2\nn\\[2mm]
 &\hspace{2.2cm}- \left(\frac{632}{27}-\frac{40}{27}\pi^2 +
\frac{224}{9}L_\mu + \frac{128}{9}L_\mu^2\right)N_L\nn\\[2mm]
 &\hspace{2.2cm} +
\left(\frac{4}{27}\pi^2+\frac{112}{9}L_\mu+\frac{160}{9}L_\mu^2\right)N_H\Bigg\}\nn\\[2mm]
 &\qquad
-\frac{99004}{27}+\frac{4954}{27}\pi^2-\left(\frac{154288}{27}-\frac{7184}{27}\pi^2\right)L_\mu
   -\frac{31232}{9}L_\mu^2-\frac{47872}{27}L_\mu^3\nn\\[2mm]
 &\qquad + \left(\frac{728}{9}-\frac{140}{27}\pi^2
+\left(\frac{2528}{27}-\frac{160}{27}\pi^2\right)L_\mu +
    \frac{448}{9}L_\mu^2 + \frac{512}{27}L_\mu^3 -
\frac{512}{27}\zeta_3\right)N_L\nn\\[2mm]
 &\qquad -\left(\frac{14}{27}\pi^2 +
\left(\frac{1264}{27}-\frac{56}{27}\pi^2\right) L_\mu + \frac{560}{9}
L_\mu^2 +
    \frac{1216}{27}L_\mu^3 - \frac{16}{27}\zeta_3\right)N_H +
\frac{19072}{27}\zeta_3\nn
\end{align}

\begin{align}
X_{42}^{\rm bare} &= \left(\frac{\mu}{m_b}\right)^{6 \epsilon} \Bigg\{
\frac{7.55556}{\epsilon^4}
-\frac{0.444444 \, N_L - 52.6667}{\epsilon^3} -\frac{3.40741 \, N_L - 188.729}{\epsilon^2} \nn\\[2mm]
&\qquad -\frac{13.2072 \, N_L - 557.390}{\epsilon} -40.3041 \, N_L + 1877.18 \Bigg\}
 \nn\\[4mm]
 X_{42}^{\rm ct} &= 0\,.
\label{xresults}
\end{align}
These results are valid for $N_F$ quark flavors out of which $N_L$ are
massless, and $N_H=N_F-N_L$ have masses equal to $m_b$, i.e.~we assume that
all quark masses except $m_b$ can be neglected. We will
address this issue once more at the end of this section.

We note that we have numerically calculated the bare terms 
of order $\alpha_s^2$ with {\tt Maple} \cite{maple}, 
requiring ten digits precision. However, when writing the
corresponding results in (\ref{xresults}), 
we retained only 6 significant digits.\footnote{Actually, we also obtained 
the results for the bare
terms proportional to $\alpha_s^2 \, N_L$ in analytical form. However,
to get a uniform presentation of the results, we give these terms
in numerical form.}.

When summing all up, the poles in $\epsilon$ cancel and we obtain the final
result
\begin{align}
G_{77}(0,\mu) &= 1 + \frac{\alpha_s(\mu)}{4\pi}
\left\{\frac{64}{9}-\frac{16}{9}\pi^2-
\frac{16}{3} L_\mu \right\} \nn\\[2mm]
&\qquad 
+ \left(\frac{\alpha_s(\mu)}{4\pi}\right)^2 \Bigg\{
(3.55556 N_F - 44.4446) L_\mu^2
  + (21.6168 N_F - 334.803) L_\mu \nn\\[2mm]
&\hspace{4cm}
+ 37.8172 N_L - 2.16077 N_H - 519.250 \Bigg\} \, ,
\label{finres}
\end{align}
noting that all the 6 significant digits given in this equation 
are in agreement with the numerical evaluation of the analytical
result for $G_{77}(0,\mu)$ 
given in ref. \cite{Blokland:2005uk}.

We also note that the terms of order $\alpha_s^2 \, N_L$ were calculated
some time ago in ref. \cite{Bieri}, using a different regularization scheme
for infrared and collinear singularities. Taking into account that in 
\cite{Bieri}
the running mass $m_b(\mu)$ appearing in (\ref{Gamma_decomp}) was 
expressed in terms of the the pole mass $m_b$, we find perfect agreement. 

Blokland et al. \cite{Blokland:2005uk} already investigated in detail the size of the
$O(\alpha_s^2)$ corrections to the ratio 
$R=\Gamma(b \to X_s \gamma)/\Gamma(b \to X_u e \bar{\nu})$ 
and the goodness of the naive non-abelianization hypothesis
in various conventions for the factor of $m_b$ that
normalize the decay rate. 
As we completely agree with their findings, we do not discuss this issue again. 
 
Instead, we briefly investigate the sensitivity of the order
$\alpha_s^2$ corrections (of course only those related to the $(O_7,O_7)$ interference)
on the charm quark mass. As discussed above, the results in
(\ref{finres}) are valid for $N_H$ flavors with mass $m_b$ and for
$N_L$ massless flavors. As the charm quark mass is somewhere in between,
it is interesting to consider the coefficient of $(\alpha_s/(4\pi))^2$ in (\ref{finres}) 
for the two extreme cases $(N_H=1,N_L=4)$ and $(N_H=2,N_L=3)$. Denoting this
coefficient by $X_2(N_H,N_L)$, we obtain (for $\mu=m_b$)
\be
X_2(1,4)=-370.142 \qquad 
X_2(2,3)=-410.120 \, . 
\ee
The true answer is expected to be somewhere in between.
As a consequence the theoretical uncertainty of this coefficient
due to the uncalculated
$m_c$-dependence is about $\pm 5\%$.

\section{Details of the calculation}

As already discussed in the introduction, we do not calculate  the
self-interference of the operator ${\cal O}_7$ by taking the
imaginary parts of the corresponding 3-loop diagrams via the optical
theorem,
as done in \cite{Blokland:2005uk}, but we directly work out the individual
two-, three- and four-particle cut-contributions as indicated in
(\ref{twothreefour}).

In a first subsection we briefly summarize the main practical tools used for
the
derivation of the unrenormalized contributions to these cuts
(denoted by $X_{ij}^{\rm{bare}}$ in section \ref{sec:2}).
In a second subsection we illustrate that we can  
obtain to very high numerical precision the results for specific diagrams in 
\cite{Blokland:2005uk} 
when summing the contributions of the corresponding cuts.
The relevant renormalization constants for the various fields and
parameters
which are needed to calculate the counterterm contributions
can be found in the third subsection.

\subsection{Regularized bare contributions} 

The starting point is the general expression for the decay rate of the
massive $b$
quark with momentum $p_b$ into $2\le n\le4$ massless final-state particles
with
momenta $k_i$,
\begin{align}\label{general_phasespace}
  \Gamma_{1\to n} &= \frac{1}{2m_b}
\left(\prod_{i=1}^n\int\!\frac{d^{d-1}k_i}{(2\pi)^{d-1}2E_i}\right)(2\pi)^d
    \,\delta^{(d)}\left(p_b-\sum_{i=1}^nk_i\right)\,|M_n|^2\nn\\[1mm]
  &= \frac{1}{2m_b}(2\pi)^n
    \left(\prod_{i=1}^{n-1}\int\!\frac{d^dk_i}{(2\pi)^d}\,\delta(k_i^2)\,
      \theta(k_i^0)\right) \nn\\[1mm]
  &\qquad\times\delta\left(\left(p_b-\sum_{i=1}^{n-1}k_i\right)^2\right)\,
    \theta\left(p_b^0-\sum_{i=1}^{n-1}k_i^0\right)\,|M_n|^2\,,
\end{align}
where the squared Feynman amplitude $|M_n|^2$ is always understood to be
summed
over final spin-, polarization- and color states, and averaged over the
spin
directions and colors of the decaying $b$ quark. For $n=4$ it also
includes a possible
factor of 1/2 if there are two identical particles in the final state.
Furthermore, $d=4-2\epsilon$ denotes the space-time dimension that we use
to
regulate the ultraviolet, infrared and collinear singularities.

\begin{figure}[t]
\linethickness{0.5mm}
\begin{picture}(450,80)(0,0)
  \put(30,40){\line(1,0){30}}
  \put(90,40){\line(1,0){30}}
  \DashLine(75,5)(75,75){2}
  \PhotonArc(75,40)(15,280,0){1}{6}
  \PhotonArc(75,40)(15,180,260){1}{6}
  \CArc(75,40)(15,0,80)
  \CArc(75,40)(15,100,180)
  \GlueArc(55,40)(10,48,180){1}{6}
  \GlueArc(55,40)(20,45,180){1}{12}
  \put(180,40){\line(1,0){30}}
  \put(240,40){\line(1,0){30}}
  \DashLine(225,5)(225,75){2}
  \PhotonArc(225,40)(15,280,0){1}{6}
  \PhotonArc(225,40)(15,180,260){1}{6}
  \CArc(225,40)(15,0,80)
  \CArc(225,40)(15,100,180)
  \GlueArc(210,40)(15,60,180){1}{6}
  \GlueArc(215,40)(30,75,180){1}{12}
  \GlueArc(220,50)(18,-5,68){1}{5}
  \put(330,40){\line(1,0){30}}
  \put(390,40){\line(1,0){30}}
  \DashLine(375,5)(375,75){2}
  \PhotonArc(375,40)(15,280,0){1}{6}
  \PhotonArc(375,40)(15,180,260){1}{6}
  \CArc(375,40)(15,0,80)
  \CArc(375,40)(15,100,180)
  \GlueArc(375,55)(12,205,260){1}{3}
  \GlueArc(375,55)(12,280,335){1}{3}
  \GlueArc(375,40)(25,0,85){1}{8}
  \GlueArc(375,40)(25,95,180){1}{8}
\end{picture}
\caption{\sf Sample contribution to $\Gamma_{1\to n}$ for $n=2$ (left),
$n=3$
  (center) and $n=4$ (right) at $O(\alpha_s^2)$. The vertical line
separates the
  original Feyman diagrams which enter the squared amplitude $|M_n|^2$.
  Thick lines denote the massive $b$ quark, thin lines the massless
  $s$ quark, wavy lines the photon, and curly lines gluons.
}
\label{somediags}
\end{figure}
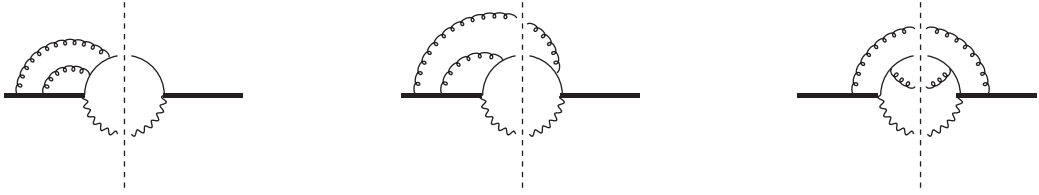

At this point a remark concerning the polarization sums over transverse
gluons is in order.
There are two well known procedures: a) One takes the full
expression for the transversal polarization sum, i.e
\be\label{pol_sum}
\sum_{r=1}^2\epsilon_{r\mu}\epsilon_{r\nu} = -g_{\mu\nu} +
\frac{t_\mu k_\nu+k_\mu t_\nu}{k\cdot t} - \frac{t^2k_\mu k_\nu}{(k\cdot
t)^2}\,,
\ee
where the polarization vectors $\epsilon_r$, the momentum $k$ of the
gluon, and the arbitrary vector $t$ are chosen such that
\be
\epsilon_r\cdot k = 0,\qquad \epsilon_r\cdot \epsilon_{r'} =
-\delta_{rr'},\qquad
\epsilon_r\cdot t = 0,\qquad r,r' = 1,2 \,.
\ee
b) Alternatively, one replaces the polarization sum by $-g_{\mu \nu}$
and includes also (unphysical) ghost particles in the final state.
We have checked that the two  methods give the same results.

The next step concerns the algebraic reduction of the quantities
$\Gamma_{1\to n}$
being composed of several contributions one of which is depicted
in Fig.~\ref{somediags} for the cases $n=2$, 3 and 4 at
$O(\alpha_s^2)$. Following \cite{Anastasiou:2002yz,Anastasiou:2003ds}, we
replace the delta functions in the phase-space representation
(\ref{general_phasespace}) by the differences of propagators, e.g.,
\be\label{deltoprops}
\delta\!\left(q^2\right) = \frac{1}{2\pi i}
\left(\frac{1}{q^2-i0}-\frac{1}{q^2+i0}\right)\, ,
\ee
in order to convert the phase space integrations into loop integrations
(which can be
combined with possible loop integrations already present in $|M_n|^2$).
In this way the systematic Laporta algorithm \cite{Laporta:2001dd} based
on the integration-by-part technique first proposed in
\cite{Tkachov:1981wb,Chetyrkin:1981qh} can be applied to reduce
scalar products in the
numerators and the number of propagators\footnote{Additional kinematical
constraints on the phase space integrations have to be implemented via
additional delta-functions
in order to be compatible with the integration-by-parts equations
\cite{Anastasiou:2003ds}.
For example the implementation of a lower cut-off $E_0$ in the photon
energy
spectrum can be achieved by inserting the term
\be
 \int_{E_0}^{\frac{m_b}{2}} dE_\gamma\,\delta
 \left(E_\gamma-\frac{p_b\cdot p_\gamma}{m_b}\right)\nn
\ee
on the right hand side of (\ref{general_phasespace}).}.
At $O(\alpha_s^2)$ this means that
we have to reduce 3-loop diagrams, see Fig.~\ref{somediags}, for which we
use
the AIR implementation of the Laporta algorithm \cite{Anastasiou:2004vj}
and a
self-written one in {\tt Mathematica} \cite{mathematica}.

After the reduction process it usually happens that in some terms the
propagators
introduced via (\ref{deltoprops}) appear with zero or negative power. In
this
case the $\pm i0$ prescription becomes irrelevant and as a consequence
these
terms cancel. In the remaining terms we convert
the propagators introduced via (\ref{deltoprops}) back to
delta-functions.

The final task is thus to perform possible loop
integrations contained in $|M_n|^2$ together with the phase space
integrations.
For $n=2$ the phase space integration is trivial,
\be
 \Gamma_{1\to2}= \frac{1}{4}
 \frac{\Gamma\left(\frac{d-2}{2}\right)}{
   \Gamma(d-2)}\,(4\pi)^{1-\frac{d}{2}}m_b^{d-5}|M_2|^2
\ee
and the loop integrations at $O(\alpha_s^2)$ can be done numerically by
means of
the sector decomposition method \cite{Binoth:2003ak}. The remaining loop
integrations at $O(\alpha_s)$ contained in $|M_3|^2$ can be performed
via introduction of Feynman parameters as usual. The integration of the
latter will also be done numerically by means of
the sector decomposition method when combined with an
appropriate parametrization of the three-particle phase space, 
see below.
For the treatment of the $n=4$ contributions we follow 
\cite{Anastasiou:2003gr} as far as the phase space representations
are concerned (see also \cite{Gehrmann-DeRidder:2003bm}).
However, as the structure of the matrix elements in our
application is slightly different than in \cite{Anastasiou:2003gr},
certain substitutions of the integration variables are necessary as described 
below. The (numerical) evaluation of the phase space integrals is then 
again performed by sector decomposition techniques.

To make the paper self-contained as far as parametrizations of
3- and 4-particle phase spaces are concerned, 
we look at them in turn.
To this end, we define the dimensionless Mandelstam variables
\be
y_{ij\dots k} = \frac{1}{m_b^2}s_{ij\dots k} =
\frac{1}{m_b^2}(k_i+k_j+\dots+k_k)^2 \, .
\ee
For the three-particle process $b\to s\gamma g$ we obtain
\be
  \Gamma_{1\to3} =
    \frac{1}{4}\frac{(4\pi)^{1-d}}{\Gamma(d-2)}\,m_b^{2d-7}
  \int_0^1\!dx_1dx_2\,(1-x_1)
    [x_1(1-x_1)^2x_2(1-x_2)]^\frac{d-4}{2}|M_3|^2\,,
\label{threeparticle}
\ee
where the variables $y_{ij}$ appearing
in $|M_3|^2$ are understood to be expressed in terms of the two
integration variables,
\be
 y_{12}=(1-x_1)(1-x_2),\qquad y_{23}=(1-x_1)\,x_2,
 \qquad y_{13}=x_1\,.
\ee
For the four-particle processes $b\to s\gamma gg$ and $b\to s\gamma
q\bar q$ we take as our starting point (see \cite{Anastasiou:2003gr})
\begin{align}
  \Gamma_{1\to4} &= (4\pi)^{-\frac{3d}{2}}
    m_b^{3d-9}\frac{2^{2d-8}\Gamma(\frac{d-2}{2})}{(d-3)\Gamma(d-3)^2}
   \int_0^1\!dx_1dx_2dx_3dx_4dx_5[x_1(1-x_1)]^{d-3}\nn\\[2mm]
 &\qquad\times x_2^\frac{d-4}{2}(1-x_2)^{d-3}
  [x_3(1-x_3)x_4(1-x_4)]^\frac{d-4}{2}[x_5(1-x_5)]^\frac{d-5}{2}|M_4|^2\,,
\label{fourparticle}
\end{align}
where the variables $y_{ij}$ appearing in
$|M_4|^2$
are understood to be expressed in terms of the five integration variables,
\begin{align}
  y_{12} &= (1-x_1)(1-x_2)(1-x_3)\nn\\[1mm]
  y_{13} &= (y_{13}^+-y_{13}^-)x_5+y_{13}^-\nn\\[1mm]
  y_{14} &= (1-x_1)(x_2+x_3-x_2x_3)-(y_{13}^+-y_{13}^-)x_5-y_{13}^-\nn\\[1mm]
  y_{23} &= x_1(1-x_2)x_4\nn\\[1mm]
  y_{24} &= x_1(1-x_2)(1-x_4)\nn\\[1mm]
  y_{34} &= x_1x_2\nn\\[1mm]
  y_{13}^\pm &= (1-x_1)\left[x_3x_4+x_2(1-x_3)(1-x_4)\pm
    2\sqrt{x_2x_3(1-x_3)x_4(1-x_4)}\,\right]\,.
\end{align}

\noindent Of course this is not the only possible parametrization since the pure
phase
space (\ref{general_phasespace}) with $|M_4|^2=1$ is symmetric with
respect
to interchanges of the momenta $k_i$. Indeed, this symmetry sometimes can
be
used to avoid the appearance of the square roots $y_{13}^\pm$ in the
squared
amplitude. We remark that the formulas given above for the three-
and the
four-particle phase spaces have the crucial property
that the pure phase-space integrals completely factorize when written in
these variables.

The expression in (\ref{fourparticle}) is a convenient 
parametrization for the $1 \to 4$ decay width for the cases where the 
denominators in $|M_4|^2$ do not depend on $y_{13}$ and $y_{14}$. As
in \cite{Anastasiou:2003gr} one can always remap the momenta of the
final state particles in a given diagram such that $y_{14}$ does not occur.
In our cases, however, structures of the form $(a-y_{13})$ appear
in the denominator, where $a$ is a linear combination of some of the 
remaining $y_{ij}$ such that $(a-y_{13})>0$. 
This is in contrast to the applications considered in   
\cite{Anastasiou:2003gr} where only $y_{13}$ appears in the denominators. 
For this reason 
we generalize the expressions given in eqs. (32) and (33) of 
\cite{Anastasiou:2003gr} to our case.
Substituting $x_5$ in (\ref{fourparticle}) by $\hat{x}_5$, where
\be
\hat{x}_5 = \frac{y_{13}(x_5) - y_{13}^{-}}{y_{13}^{+} - y_{13}^{-}} \,
\frac{a-y_{13}^+}{a-y_{13}(x_5)} \, ,
\ee
we obtain the representation
\begin{align}
  \Gamma_{1\to4} &= (4\pi)^{-\frac{3d}{2}}
    m_b^{3d-9}\frac{2^{2d-8}\Gamma(\frac{d-2}{2})}{(d-3)\Gamma(d-3)^2}
   \int_0^1\!dx_1dx_2dx_3dx_4d\hat{x}_5[x_1(1-x_1)]^{d-3}\nn\\[1mm]
 &\qquad\times x_2^\frac{d-4}{2}(1-x_2)^{d-3}
  [x_3(1-x_3)x_4(1-x_4)]^\frac{d-4}{2}[\hat{x}_5(1-\hat{x}_5)]^\frac{d-5}{2}
  (a-y_{13}(\hat{x}_5))\nn\\[3mm]
 &\qquad\times [(a-y_{13}^+)(a-y_{13}^-)]^\frac{d-5}{2} 
\left[ (1-\hat{x}_5)(y_{13}^{-} - y_{13}^{+})+(a-y_{13}^{-}) \right]^{4-d} \, 
|M_4|^2 \, ,
\label{fourparticlehat}
\end{align}
in which the factors of $(a-y_{13})$ that appear in the denominators
of $|M_4|^2$ get cancelled.

\subsection{Illustrative example}

\begin{figure}[t]
\linethickness{0.5mm}
\begin{picture}(450,80)(0,0)
  \put(180,40){\line(1,0){30}}
  \put(240,40){\line(1,0){30}}
  \DashLine(230,5)(230,75){2}
  \DashLine(205,10)(265,70){2}
  \DashLine(185,70)(262,10){2}
  \PhotonArc(225,40)(15,180,0){1}{12}
  \CArc(225,40)(15,0,180)
  \GlueArc(210,45)(18,35,195){1}{12}
  \GlueArc(220,45)(18,10,195){1}{12}
\end{picture}
\caption{\sf Three-loop diagram $(o)$ of reference \cite{Blokland:2005uk}. 
The dashed lines indicate the 2-,3- and 4-particle cuts (see text).
Thick lines denote the massive $b$ quark, thin lines the massless
$s$ quark, wavy lines the photon, and curly lines gluons.
}
\label{example}
\end{figure}
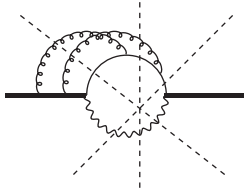

As mentioned, Blokland et al. \cite{Blokland:2005uk}
calculated the order $\alpha_s^2$ contributions to the decay width by working
out (according to the optical theorem) the absorptive part of the $b$-quark 
self-energy. At this order the (bare) self-energy is given in terms of
the 3-loop diagrams depicted in Fig. 1 of \cite{Blokland:2005uk}.
As an example we consider their diagram $(o)$ and redraw it 
in Fig.~\ref{example}. The diagram has a two-, a three- and a four-particle
cut as indicated by the straight lines. These cuts represent interferences
which are contained in the quantities $\Gamma_{1 \to n}\;(n=2,3,4)$ defined above. 
We denote these interferences by $\Gamma^{(o)}_{1 \to n}$, where the
superscript stands for diagram $(o)$. As 
we explained in detail in the previous subsection how to calculate such
interferences,
we immediately give the result (putting $\mu=m_b$ and working in the 
Feynman gauge): 
\be
\Gamma^{(o)}_{1 \to n}=\Gamma_0 \left( \frac{\alpha_s^2}{4\pi}\right)^2 \, 
C_F \left(C_F-\frac{C_A}{2}\right) \, 
\left( Y_2^{(o)} + Y_3^{(o)} + Y_4^{(o)} \right) \, ,
\label{diago_decomp}
\ee
where $\Gamma_0$ denotes the decay width at lowest order in QCD.
The quantities $Y_2^{(o)}$, $Y_3^{(o)}$, $Y_4^{(o)}$, corresponding to
the 2- 3- and 4-particle cut, read:
\begin{eqnarray}
Y_2^{(o)}&=&0.333333/\epsilon^4
        + 1.66667/\epsilon^3 
        + 2.08693/\epsilon^2   
        - 6.64830/\epsilon 
        - 67.1138 \nn \\
Y_3^{(o)}&=&-1.00000/\epsilon^4
          -5.00000/\epsilon^3 
         - 0.970923/\epsilon^2   
        +  72.0762/\epsilon 
        + 410.481 \nn \\
Y_4^{(o)}&=&0.666667/\epsilon^4
          +3.33333/\epsilon^3 
         -  1.11601/\epsilon^2   
        -  61.4279/\epsilon 
       -316.587  \, .
\end{eqnarray}
For the sum $Y^{(o)}=Y_2^{(o)}+Y_3^{(o)}+Y_4^{(o)}$ we obtain
\be
Y^{(o)}=  4.00000/\epsilon + 26.7798   \, ,
\ee
which is in perfect agreement with the analytic
result
\be
Y^{(o)}= 4/\epsilon +\left[\frac{545}{6}-\frac{50}{3}\zeta(3)+\frac{28}{3}\pi^2 \ln
2 -14\pi^2+\frac{14}{45}\pi^4\right] = 4.00000/\epsilon +26.7798
\ee
of Blokland et al. \cite{Blokland:2005uk}.

\subsection{Renormalization constants}

In this subsection, we collect the relevant renormalization constants
necessary to render all bare amplitudes ultraviolet finite and fix the adopted
renormalization schemes.

\noindent First, the strong coupling constant is renormalized using the $\MS$ scheme:
\be\label{alpha_ms}
 \alpha_s^{\rm bare} =
 \mu^{2\epsilon}\left(\frac{e^\gamma}{4\pi}\right)^\epsilon
 Z_\alpha^{\MS}\,\alpha_s(\mu)\,,
\ee
with
\be
 Z_\alpha^{\MS} = 1 - \frac{33-2N_F}{3\,\epsilon}\frac{\alpha_s(\mu)}{4\pi} +
 O(\alpha_s^2)\,.
\ee

Also the $b$ quark mass contained in the operator $O_7$ and the Wilson
coefficient $C_7^{\rm eff}$ are renormalized in the $\MS$ scheme in our calculation. 
In the present application we only need
the product of these two renormalization constants for which we find
\be
 Z_{m_b}^{\MS}Z_{77}^{\MS} = 1 +
 \frac{4}{3\,\epsilon}\frac{\alpha_s(\mu)}{4\pi} -
 \left(\frac{58-4N_F}{9\,\epsilon^2}-\frac{543-26N_F}{27\,\epsilon}\right)
 \left(\frac{\alpha_s(\mu)}{4\pi}\right)^2 + O(\alpha_s^3)\,,
\ee
in full agreement with \cite{Misiak:1994zw,Gambino:2003zm}. 

All the remaining fields and parameters are 
renormalized in the on-shell scheme. The on-shell renormalization constant for
the $b$ quark mass is given by
\be
 Z_{m_b}^{\rm OS} =
 1-\frac{4}{3}\,\Gamma(\epsilon)\,e^{\gamma\epsilon}\,
 \frac{3-2\epsilon}{1-2\epsilon}
 \left(\frac{\mu}{m_b}\right)^{2\epsilon}\frac{\alpha_s(\mu)}{4\pi} +
 O(\alpha_s^2)\, .
\ee
For those of the gluon-field and the $s$ quark-field we find  
\begin{align}
  Z_3^{\rm OS} &=
    1-N_H\frac{2}{3}\,\Gamma(\epsilon)\,e^{\gamma\epsilon}
    \left(\frac{\mu}{m_b}\right)^{2\epsilon}
    \frac{\alpha_s(\mu)}{4\pi} + O(\alpha_s^2)\nn\\[2mm]
  Z_{2s}^{\rm OS} &= 1 + N_H\frac{4}{3}
    \frac{\epsilon\,(1+\epsilon)(3-2\epsilon)
      \Gamma(\epsilon)^2\,e^{2\gamma\epsilon}}
    {(1-\epsilon)(2-\epsilon)(1+2\epsilon)(3+2\epsilon)}
    \left(\frac{\mu}{m_b}\right)^{4\epsilon}
    \left(\frac{\alpha_s(\mu)}{4\pi}\right)^2
    + O(\alpha_s^3)\,.
\end{align}

Finally, the field renormalization constant of the $b$ quark in the
on-shell scheme can be found in the literature \cite{Melnikov:2000zc}:
\begin{align}
  Z_{2b}^{\rm OS} &=
    1-\frac{4}{3}\,\Gamma(\epsilon)\,e^{\gamma\epsilon}\,
    \frac{3-2\epsilon}{1-2\epsilon}
\left(\frac{\mu}{m_b}\right)^{2\epsilon}\frac{\alpha_s(\mu)}{4\pi}\nn\\[2mm]
  &\quad + \Bigg\{\frac{1}{\epsilon^2}\left(30-\frac{4}{3}N_L\right) +
    \frac{1}{\epsilon}\Bigg[\frac{22}{9}N_L +
    \left(\frac{2}{3}+\frac{16}{3}L_\mu\right)N_H\nn\\[2mm]
  &\hspace{15mm} -\frac{59}{3}
    +32L_\mu\Bigg]+\left(\frac{113}{9}+\frac{8}{9}\pi^2 +
    \frac{152}{9}L_\mu + \frac{16}{3}L_\mu^2\right)N_L\nn\\[2mm]
  &\hspace{15mm}+\left(\frac{947}{27}-\frac{10}{3}\pi^2+\frac{88}{9}L_\mu
    +16L_\mu^2\right)N_H-\frac{3383}{18}-\frac{16}{9}\pi^2\nn\\[2mm]
  &\hspace{15mm} -\frac{32}{9}\pi^2\ln
    2+\frac{16}{3}\zeta_3-196L_\mu -
    24L_\mu^2\Bigg\}\left(\frac{\alpha_s(\mu)}{4\pi}\right)^2 +
O(\alpha_s^3)\,.
\end{align}

\section{Conclusions}

We have presented a calculation of the $O(\alpha_s^2)$ contribution associated
with the electromagnetic dipole operator ${\cal O}_{7}$ to the decay width for 
$\bar B \rightarrow X_s \gamma$.
In contrast to a previous work \cite{Blokland:2005uk},
the calculational method we used is straightforwardly
extendable to the calculation of the interference of the electromagnetic and 
chromodynamical operator and also to the self-interference of the latter.
 Our final results are in complete agreement with those in 
\cite{Blokland:2005uk}. Additionally, we estimated that the $m_c$-dependence 
of the $O(\alpha_s^2)$ contribution of the 
$({\cal O}_7,{\cal O}_7)$-interference, 
which is not worked out in both papers, leads to an uncertainty of about $5\%$
(of this $O(\alpha_s^2)$ contribution).

The NNLL piece calculated in the present paper is an 
important ingredient of the complete NNLL 
prediction for the decay rate of the process $\bar B \rightarrow X_s \gamma$.
According to a previous study, it is expected that the theoretical error 
of the forthcoming NNLL result will be reduced by factor $2$ \cite{New}. 
This increases the high sensitivity of the decay 
$\bar B \rightarrow X_s \gamma$ to possible new degrees of 
freedom even further. 

\section*{\normalsize Acknowledgements}
\vspace*{-2mm}

The team from Yerevan is partially supported by the
ANSEF N 05-PS-hepth-0/25-338 program.
C.G. and T.E. are supported by the Swiss National Foundation; RTN,
BBW-Contract No.01.0357 and EC-Contract HPRN-CT-2002-00311 (EURIDICE).

\end{document}